# The Synthesis of Regression Slopes in Meta-Analysis

**Betsy Jane Becker and Meng-Jia Wu**

*Abstract.* Research on methods of meta-analysis (the synthesis of related study results) has dealt with many simple study indices, but less attention has been paid to the issue of summarizing regression slopes. In part this is because of the many complications that arise when real sets of regression models are accumulated. We outline the complexities involved in synthesizing slopes, describe existing methods of analysis and present a multivariate generalized least squares approach to the synthesis of regression slopes.

We begin with a discussion of the rationale for summarizing regression slopes, a practice that has become more prevalent in meta-analyses in recent years. We then examine the methods for summarizing slopes that have been proposed to date, and the assumptions and data requirements of those methods. We conclude by presenting a generalized least squares (GLS) approach to the synthesis of regression slopes for continuous predictors and outcomes, with remarks on the challenges and limitations to synthesis of such estimates.

## 1. SYNTHESIZING SLOPES

While it is by no means common or well understood, the synthesis of regression slopes has received increased attention in recent years (e.g., Baker et al. (2003); Peterson and Brown (2005); Roberts (2005);

*Betsy Jane Becker is Professor of Measurement and Statistics, College of Education, Florida State University, Tallahassee, Florida 32306, USA e-mail: bbecker@fsu.edu. Meng-Jia Wu is Assistant Professor of Research Methodology, School of Education, Loyola University Chicago, Chicago, Illinois 60611, USA e-mail: mwu2@luc.edu.*

Rose and Stanley (2005)). This growing interest is likely related to the increasingly complex models investigated in primary research, at least in the social sciences. Researchers want to model the effects of multiple predictors as well as to control for potential confounding variables, and in the context of a primary study this is often achieved by including such variables in complex models. Results of techniques like structural equation modeling, hierarchical linear modeling and multiple regression have often been omitted from meta-analyses because of a lack of knowledge about how to synthesize indices from these analyses, and because of the complexities and assumptions underlying the process of synthesis.

The main purposes of this paper are to point out the complexities and potential problems in synthesizing slopes from regression models, to describe existing methods for summarizing slopes and to present a new synthesis approach based on generalized least squares estimation. We focus only on the case of multiple regression, though clearly other analyses involve regression-like models with similar assumptions. We begin with the simple case where all studies examine very similar models and discuss techniques for estimating a combined regression model across studies. Modeling to examine the impact of study features, design differences and study quality (e.g., Pang, Drummond and Song, 1999) is touched on briefly. Other complications such as publication bias (e.g., Doucouliagos (2005); Stanley (2005)) are beyond the scope of our discussion.

Consider a model in study $i$ relating some predictors $X_1$ through $X_P$ to an outcome $Y$ for case $j$. Specifically, in study $i$,

$$(1) \quad Y_{ij} = \beta_{i0} + \beta_{i1}X_{ij1} + \cdots + \beta_{iP}X_{ijP} + e_{ij}$$

for $j = 1$ to $n_i$ cases. The usual assumptions of normality and homoscedasticity of errors apply such that $e_{ij} \sim N(0, \sigma_i^2)$, and linearity of the $X$–$Y$ relations is also assumed within each study. Often in a synthesis one predictor (let us say $X_1$) is of primary interest; below we refer to this as the focal predictor. Now assume we have a series of studies $i = 1$ to $k$, and each of them involves a regression with $X_1$ as a predictor of $Y$; typically also other predictors (say, $X_2$ through $X_P$) appear in these studies. We may wish to summarize the slopes representing the relation of $X_1$ to $Y$ (estimates of $\beta_{11}$ through $\beta_{k1}$), and on occasion perhaps to summarize all $P$ slopes for $X_1$ through $X_P$, across the $k$ studies.

While syntheses of slopes imply a variety of fairly stringent assumptions, this has not deterred researchers from combining regression slopes (though some existing summaries have been done without regard to the underlying assumptions). Crouch (1995, 1996) summarized slopes from a diversity of models representing tourism demand and Lau and colleagues (Lau, Sigelman, Heldman and Babbitt, 1999) used regressions to examine the effectiveness of negative political advertisements. Farley, Lehmann and Sawyer (1995) encouraged marketing researchers to synthesize regression slopes, and more recently Peterson and Brown (2005) reviewed the use and synthesis of standardized slopes in meta-analysis in the field of psychology. Two controversial and very different syntheses of regression results in education dealt with the topic of whether educational expenditures relate to achievement outcomes (Hanushek (1989); Hedges, Laine and Greenwald, 1994). A recent issue of the *Journal of Economic Surveys* (e.g., Roberts (2005); Rose and Stanley (2005)) focused exclusively on meta-analyses of regression coefficients on a variety of economic topics, and many others have synthesized regressions on diverse topics in economics (e.g., Card and Krueger (1995); Doucouliagos and Paldam (2006)), in large part thanks to the seminal work of Stanley and Jarrell (1989).

In spite of their widespread use in economics, methods for summarizing regression slopes have received less attention in the statistical literature than methods for synthesizing other indices used in meta-analysis such as standardized mean differences, correlations, and proportions (or transformed proportions such as odds ratios). Analytic approaches may be proposed in the methods sections of substantive syntheses without much attention to the statistical behavior of the estimators and tests involved. In this article we provide a multivariate formulation for the synthesis of slopes, beginning with discussions of the assumptions required and of problems that meta-analysts may encounter when synthesizing slopes. We then briefly review several existing univariate and multivariate approaches. Most existing approaches are univariate, which avoid some, but not all, of the issues and assumptions that underlie the synthesis of sets of regression slopes. Other approaches to combining slopes are more complex, but require access to raw data which is quite unusual to have in the meta-analysis context.

## 2. ASSUMPTIONS AND PROBLEMS IN SYNTHESIZING SLOPES

The synthesis of regression slopes is difficult for several reasons, and a variety of problems must be dealt with in the process. Problems include nonequivalence of the metrics for the predictors and outcomes across studies, lack of information in study reports and the estimation of very diverse models across studies. Slopes are identically distributed across studies when the outcome $Y$ and the focal predictor $X$ are measured similarly, when the same additional $X$s appear in each study (i.e., the same model is estimated in each study), and when $X$ and $Y$ scores are similarly distributed. Each of these conditions is often not met across studies, which is a concern for the meta-analysis of slopes. We consider each condition in turn.

### 2.1 $Y$ Is Measured Similarly Across Studies

This consideration is important because even if only a single predictor appears in each of a collection of $k$ regression equations, the raw regression slope in each study depends on the scales of that predictor and the outcome. This is evident in the language commonly used to describe the raw regression slope—the predicted change in the outcome $Y$ given one unit change in $X$. Also this is easily seen in the formula for the slope in a bivariate regression, which is $b = r_{XY}(S_Y/S_X)$. Here $r_{XY}$ is the correlation between $X$ and $Y$, $S_Y$ is the standard deviation of the $Y$ scores and $S_X$ is the standard deviation of the $X$ scores. For two raw-scale slopes to be comparable across studies, the scales of $Y$ and $X$ must be the same (or proportional, e.g., both $X$ and $Y$ could be linearly transformed using the same transformation). Indeed, for total equivalence of scales, the measures of $Y$ and $X$ should be equally reliable across studies, which is rare (Amemiya and Fuller (1984); Hunter and Schmidt (2004)).

We consider an example where $Y$ is a measure of the quality of teaching—often represented as student achievement. Our examples are drawn from an ongoing synthesis of studies of the relationship of teacher qualifications to measures of the quality of teaching. Across studies student achievement is typically measured using different tests of different constructs (math, reading, etc.), which may be presented as posttests, difference scores or other measures of change over time and the like. We have identified over 190 studies that examine measures of student learning and to date, from 65 studies with measures coded in detail, we have identified 79 different measures of student learning (and coding is not complete). At least in this synthesis, $Y$ is *not* measured similarly across studies.

In some areas, particularly in economics where outputs may be monetary, outcomes will be measured similarly or can be transformed or adjusted to be reasonably similar. For instance, Ashenfelter, Harmon and Oosterbeek (1999) examined returns to schooling, where the outcome was earnings, and earnings scales can be reasonably well equated across countries and over time. However, in many areas this will not be possible.

### 2.2 Focal $X$ Is Measured Similarly Across Studies

This is also a problematic assumption. Again in some realms, such as the study of economic inputs measured in dollars or other forms of currency, this may not be an issue (e.g., per pupil expenditures were examined by Hedges, Laine, and Greenwald (1994)). In the Ashenfelter, Harmon and Oosterbeek (1999) review, schooling was apparently measured in years, which would also be comparable across studies. Even in such cases, however, adjustments (e.g., for inflation, for exchange rates) will sometimes be required. Also when the index of study results is an elasticity (common in economics) and represents proportional change in $X$ and $Y$, the scale of $X$ may not be as critical. In other areas, however, the focal $X$'s may not be measured similarly. Shi and Copas (2004) noted that exposure (dose) variables are often measured categorically in medical dose-response studies. They referred to the problem of having such categorizations (which can vary across studies) as the problem of "grouped dose levels."

In our synthesis of the literature on teacher qualifications, studies examine such predictors as degrees earned, counts of courses taken, numbers of credits taken, performance on teacher tests and teaching experience. Some of these (e.g., counts of courses taken) may be measured fairly similarly across most studies, while others (teacher test performance) are not. Even such things as teaching experience are not always measured as ratio-scale, continuous variables (e.g., years of experience). We have found such variations as dichotomies representing novice versus experienced teachers, categorical representations (e.g., teachers with 0–5 years, 6–10 years, or more than 10 years experience) and years transformed to represent nonlinear effects (e.g., squared years of experience).

### 2.3 Same Additional $X$'s Across Studies

This condition is virtually never met. In practice, studies nearly always estimate different models. In fact it can be argued that differences in the models analyzed should be expected across studies, as researchers develop and elaborate on models present in the literature in attempts to refine prediction and to explain additional variation in the outcomes of interest. Stanley and Jarrell (1989) raised a concern about differences in models in the context of model specification, and argued that syntheses of regression results should examine aspects of models such as the functional form of the variables involved and differences in the independent variables included in the regressions. Many economists have dealt with this issue by modeling regression slopes or other indices of effect as functions of dummy variable predictors that represent differences in model specification (e.g., Doucouliagos and Paldam (2006); Stanley (2001)).

Some examples of how models can vary widely come from the literature on teacher qualifications and the quality of teaching. Wu and Becker (2004) examined regression models of the impact of teacher experience on student outcomes based on two large-scale survey data sets: the Coleman Equality of Educational Opportunity (EEO) data (Coleman et al. (1966)) and the National Education Longitudinal

Study: 1988 (NELS:88) (e.g., Ingels et al. (1992)). Wu and Becker found 92 different models for the prediction of student achievement in 12 studies using the EEO data set. Nine of those had examined teacher experience. Similarly, 55 different models appeared in the 11 studies based on the NELS:88 data set; 6 of those models had examined teacher experience. More critically, other than teacher experience, the 9 models using the EEO data set together contained 122 different additional independent variables, and the 6 models based on the NELS:88 data set contained 103 other independent variables. The regression models contained a diversity of additional variables, including socioeconomic status, teacher salary, teacher/pupil ratio, school characteristics, student and family characteristics, and the like.

The question of whether models are similar across studies is important because the metric of the raw slope for $X$ depends on both the outcome $(Y)$ and $X$, and because of model specification issues. In particular, each slope's precision, degree of bias and covariation with other slopes depend on the other $X$'s in the model. To the extent that a model is not properly specified, all slopes in the model are potentially biased.

Also how intercorrelated the slopes are (i.e., the degree of multicollinearity) depends on what $X$s are included. The covariance matrix $\text{Cov}(\mathbf{b}_i)$, where $\mathbf{b}_i$ is the vector of slopes for study $i$, contains this information. [Notation and formulas for $\text{Cov}(\mathbf{b}_i)$ are introduced below.] One simple example suffices to make this point: Consider the slope for $X_1$ when there is only one additional $X$ in the model (say $X_2$). The correlation between the slopes for $X_1$ and $X_2$ [i.e., $\text{Corr}(b_{X1}, b_{X2})$] is the opposite of the bivariate correlation $\text{Corr}(X_1, X_2)$ between $X_1$ and $X_2$ (Stapleton (1995)). When additional $X$'s are added the slope covariances depend on the partial correlation between $X_1$ and $X_2$, controlling for other $X$'s. Even this simple fact reveals that each slope's distribution depends on other predictors in the model. However, in practice, the covariance matrix among the slopes in primary studies is rarely reported (though matrices of correlations among predictors are sometimes reported). So it will be unusual to find full $\text{Cov}(\mathbf{b}_i)$ matrices in published studies, and in such cases caution may be needed in synthesizing slopes from very different models.

The extent to which differences in the models estimated across studies lead to *important* differences in slopes across studies is unclear. Therefore, the question of model specification is relevant here. If all of the different versions of models are (reasonably) well specified, then each one should provide unbiased and relatively independent estimates of the regression slopes. The impact of model differences likely depends on both model specification and on the relationships of the focal $X$ to the additional variables. Let us consider the focal predictor or some "base set" of predictors that appear in a well specified model. If additional variables are relatively independent of the predictors in the base set, the slopes of the base set of predictors (and their distributions) may not be much affected by the addition of those new variables. However to the extent that added variables are highly correlated with the base predictors or with the outcome, the slopes of the base predictors will differ and will also be biased. We suspect that there will be some limitations to the application of the estimation approach shown here when the models used across different studies differ widely, and in particular when some suffer from multicollinearity or other forms of misspecification.

Some empirical investigations have attempted to shed light on the role of additional primary-study predictors on regression slopes. Peterson and Brown (2005) found no impact of either sample size or the number of additional predictors in regression models on the relation between the standardized slope and a corresponding zero-order correlation. Their analyses included slopes from an incredibly wide range of areas and encompassed different predictors and outcomes from studies in psychology, sociology, marketing and management. It is possible that by looking across so many diverse regression models the impact of the nature of the models would be diluted. Ashenfelter and colleagues (1999) attempted a more nuanced investigation in their review of studies of returns to schooling: they assessed the importance of the presence of controls for ability and measurement error in the primary-study regressions. Their analyses suggested a complicated pattern of impact of ability controls, with effects for returns to schooling in the United States increasing when ability was controlled and effects in non-U.S. studies decreasing. In contrast, the inclusion of controls for measurement error did not appear to significantly affect the slopes.

An analysis by Doucouliagos and Paldam (2006) examined models for the effects of economic development aid on the accumulation of capital in the countries that receive such aid. To explore differences in sample and model specification, their analyses examined aid-effectiveness elasticities as the outcome and included as many as 11 dummy variable predictors that represented study differences. These dummy variables represented the type of model examined in the primary study (e.g., fiscal response models versus growth equations), the nature of the data set used (its type and countries included) and the presence of three different control variables. Controls for endogeneity and the model type variables had significant impacts on the elasticities, as did sample size. In this analysis, differences in the forms of the models examined in the primary research played a large role in the synthesis results.

## 3. EXISTING METHODS FOR SUMMARIZING SLOPES

Next we examine methods that have been proposed for synthesizing regression slopes. These techniques have been described in the literature, but some do not appear to have been used in meta-analytic practice.

### 3.1 Summaries of Slopes or Functions of Slopes

Several authors have used direct and simple summaries of slopes or differences in slopes, although none has provided a clear statistical justification for the approaches used. Jarrell and Stanley (1990) used slopes for a dummy variable that represented union membership (from regression models predicting log wage values) in a review of the differences in wages between union and non-union workers. In a similar analysis, Stanley and Jarrell (1998) examined the gender gap in wages. Using ordinary least squares (OLS) regression analyses, Jarrell and Stanley examined two models for the wage gap due to union membership. One had 20 predictors representing differences in sample and model specification, and the other included 77 predictors. The initial 20 predictors represented differences in model specification such as the nature of the wage variables used and differences in the samples analyzed (e.g., whether blue-collar, white-collar or government workers were included). The other model included those 20 predictors, plus allowed those predictors to vary over time, and also included 10 indicators identifying particular data sets that were used and 27 indicator variables representing multiple findings contributed by 27 primary-study authors.

Jarrell and Stanley applied OLS in spite of acknowledging that the errors in their model were likely to be heteroscedastic, noting that "[v]arious efforts to adjust for the problem made little difference in this application" (Jarrell and Stanley (1990), page 56). Similarly Stanley and Jarrell used OLS methods and tested for homoscedasticity using "conventional tests" (Stanley and Jarrell (1998), page 961). It is not clear why these authors did not find heteroscedasticity, unless their results arose from roughly equal sized samples, because as will be shown below, slopes will typically not have equal variances across studies and thus errors from models with slopes as outcomes will typically not be homoscedastic either.

### 3.2 Summaries of $t$ Statistics

Stanley and Jarrell (1989) encouraged economists to summarize regression slopes and suggested using the $t$ statistic (i.e., the slope divided by its standard error) as an index. They suggested this metric as a way to deal with heteroscedasticity of slopes across studies, which could occur because of sample size differences and differences in precision. They also argued that dividing $b$ by its standard error removes problems due to use of different scales across studies. While summaries of $t$ values have long existed (e.g., Walker and Saw (1978)), there are some drawbacks to their use. First and of greatest concern, the $t$ contains information on sample size and precision as well as effect magnitude. Thus $t$ can become large either when the slope itself is large or when its standard error is small, which occurs both when the sample is large and when there is little variation in the regression residuals. Stanley and Jarrell argued that the $t$ "is a standardized measure of the critical parameter of interest" (2005, page 304), but they did not say what the parameter of interest is. Clearly $t$ is not an estimator of $\beta$. Also these authors do not explain whether one can use a summary of the $t$ values to obtain a slope estimate after pooling or summarizing the $t$ values. Moreover, it is sometimes difficult to determine the direction of an effect from a $t$ test if a slope is not presented and the test is not significant or when the researcher reports only the absolute value of the $t$. Given all of these concerns, $t$ values are likely to be less meaningful than other indices based on slopes when findings are to be interpreted.

While Stanley and Jarrell did not initially describe exactly how one would summarize $t$ values,

in practice what they and others have often done is to model $t$s or functions of $t$s in terms of predictors that characterize the regression models in their synthesis. For instance, Card and Krueger (1995) examined $\log |t|$ values representing the effects of different levels of the minimum wage on employment rates. They estimated ordinary least squares regressions for the $\log |t|$ values which included as predictors the log of the square root of the error degrees of freedom in the primary study, a dummy indicating whether the data included a subsample of teenagers and the number of explanatory variables in the primary-study regression model.

Based on Jarrell and Stanley's recommendation, Lau, Sigelman, Heldman and Babbitt (1999) used $t$ values in a summary of results from group comparison studies and regression studies that focused on the effect of negative political advertisements on political campaigns. They found that about one-quarter of their data points "come from ordinary least squares (OLS) or logistic regression equations, and there is no universally accepted method for handling such data in a meta-analysis" (Lau et al., 1999, page 855). To avoid losing data, they extracted $t$ statistics associated with regression coefficients that represented mean differences between groups exposed to negative advertisements and control groups (exposed to no advertisements or positive advertisements). They converted the $t$ values into standardized mean differences ($ds$), via $d = 2t/(df)^{1/2}$. They argued that the $ds$ obtained via this transformation could be combined with other $ds$ from group comparisons. However, to the extent that the primary-study regression models included other important control variables, these $t$s likely produced partial effect sizes, which do have slightly different distributions from "typical" zero-order effect sizes (Keef and Roberts (2004)).

Another index that is related to the $t$ value is Timm's (2004) "ubiquitous effect size." This index can represent a single slope or a linear combination of slopes. Timm's index resolves the problem of dependence of the $t$ on sample size because it incorporates a multiplier that reduces the influence of the sample size on its value. However, to date Timm's index has not been used in syntheses of slopes and Timm did not provide methods for synthesizing his index.

### 3.3 Iterative Least Squares Regressions

An iterative GLS approach was proposed by Hanushek (1974) to summarize slopes representing returns to schooling. Hanushek's method optimally requires the raw data in order to estimate a covariance matrix among the slopes. However, he suggested an alternative approach whereby part of the covariance matrix could be estimated using OLS regression across studies and the estimate obtained from this step would then be added to a function of raw data from the original (within-study) regressions. To the extent that the approach requires raw data and infrequently reported summary values from the original studies, it will not be applicable in many meta-analytic settings.

### 3.4 Dose-Response Models in Epidemiology

Greenland (1987), Greenland and Longnecker (1987) and Shi and Copas (2004) considered slopes that relate the amount of exposure to some substance to odds-ratio outcomes. Typical studies relate levels of exposure (e.g., to alcohol, to smoke as in passive smoking, etc.) to outcomes including diagnoses of various kinds of cancer and other diseases. These studies fit into the regression framework because researchers want to know whether the level of exposure to some substance predicts higher levels of problematic outcomes (e.g., higher rates of cancer). Some issues are similar to those for continuous outcomes, but the outcome metric differs in these cases because it is typically a dichotomy (survival versus death, presence of some disease versus no disease, etc.). In the epidemiology literature typical fixed and random-effects syntheses of the dose-response slopes have been conducted (weighting by the within-study slope variances), and the issue of dependence has been addressed by incorporating a within-study covariance between odds ratios (at different exposure levels) into the analyses. This covariance is different from the covariance between slopes, which is incorporated in our methods below.

Shi and Copas (2004) argued for the use of maximum likelihood estimators of the mean dose-response slope and a between-studies variance component for the slopes, and they also describe a likelihood test of homogeneity of the dose-response slopes. Shi and Copas considered a bivariate regression because only one predictor (exposure to the dosing variable) was used in the within-study model. They argued that their approach is also approximate for adjusted odds ratios (e.g., adjusted for age or other predictors), provided the adjustments are not great. The adjusted odds ratio case is similar to the typical situation in most areas of social science, where multiple control variables are included in each regression model.

### 3.5 Validity-Generalization Approaches

A considerable literature exists concerning the synthesis of test validities (e.g., Hunter and Schmidt (2004)). This area is known as validity generalization, with a key issue being whether test validities generalize (i.e., can be applied reasonably well) across job types and job settings. Test validities are typically indices that represent the relation between a predictive test (e.g., an employment selection test) and some later outcome such as job performance. A key issue in this literature is the effect of differential test reliability across studies, thus corrections for measurement error are a standard part of the validity-generalization approach. While test validities are most often represented by correlation coefficients, on occasion more complex regression models are used to examine test validity. Raju, Fralicx and Steinhaus (1986) estimated the mean slope and between-studies variation in slopes with corrections for unreliability in $X$, where $X$ is a predictive test whose validity is of interest. Their methods parallel those presented by Hunter and Schmidt (2004) for analyzing correlation coefficients, which have been controversial in the meta-analysis literature (see, e.g., Hedges (1988)). Even so, they were used by Crouch (1995, 1996) and Root and colleagues (2003). Later Raju, Pappas and Williams (1989) conducted an empirical Monte Carlo study of a validity data base to examine the performance of methods using slopes and correlations and covariances to represent validities.

### 3.6 Weighted Least Squares (Univariate) Approaches

The weighted least squares (WLS) approach was used by Bini, Coelho and Diniz-Filho (2001), who cited Hedges and Olkin (1985) as the basis of their approach. Greenland and Longnecker (1987) also described this approach. If we consider the model in (1) above relating some $X$s to $Y$ for person $j$ in study $i$, we may want an estimate of the slope for one predictor, say $X_1$. Estimating model (1) in each of $k$ studies (using the same estimation method, such as ordinary least squares) produces independent and approximately normally distributed estimates of the population slopes $\beta_{11}, \beta_{21}, \ldots, \beta_{k1}$. If we denote those estimates as $b_{11}, b_{21}, \ldots, b_{k1}$ we can use least squares methods to summarize the slopes. Thus, for instance, we can compute the combined slope $b_{\cdot 1}$,

$$b_{\cdot 1} = \frac{\sum_{i=1}^{k} w_{i1} b_{i1}}{\sum_{i=1}^{k} w_{i1}}, \tag{2}$$

where $k$ is the number of slopes combined, $b_{i1}$ is the slope for $X_1$ from study $i$ and $w_{i1}$ is the weight for that slope in the $i$th study, which is the reciprocal of the slope variance $[w_{i1} = 1/\text{Var}(b_{i1})]$. The variance of $b_{\cdot 1}$ is given as

$$V(b_{\cdot 1}) = \frac{1}{\sum w_{i1}}. \tag{3}$$

This approach could also be applied to partial correlations or standardized regression slopes. If standard errors were not available, one could weight by sample size, as the relevant standard errors are typically a function of $n$ or of the degrees of freedom for the regression model.

### 3.7 Multivariate Bayesian Approach

One last proposed method for simultaneously estimating a set of regression models was given by Novick and colleagues (Novick, Jackson, Thayer and Cole, 1972) in the context of the validity of college-admissions prediction, where all predictors are consistently measured across colleges. Furthering a Bayesian method attributed to Lindley, the authors argued for a multistage Bayesian formulation involving raw data, its parameters (the slopes) and hyperparameters (e.g., the variances of the slopes). However, while the method constitutes an improvement beyond the methods above because it is multivariate and uses simultaneous estimation, this approach requires full access to the raw data so is not applicable in the meta-analytic context.

## 4. MULTIVARIATE GLS APPROACH

Most of the analyses presented above are reasonable if one wants to synthesize estimates of a single population slope and if most of the studies involving that slope examine simple models. However, within the $i$th sample, the $P+1$ slopes $b_{i0}, b_{i1}, \ldots, b_{iP}$ are often correlated and there may be interest in obtaining an overall regression model (rather than a single slope estimate). To synthesize slope vectors $\mathbf{b}_1, \mathbf{b}_2, \ldots, \mathbf{b}_k$, we need generalized least squares (GLS) methods, primarily because of the unequal variances of effects for studies of different sizes. [Stanley and Jarrell argued that one could obtain estimates of the vector of slopes by solving a system of equations with the slopes as endogenous variables (Stanley and Jarrell (1989), page 169). However, they did not discuss exactly how to do so or how to deal with the fact that within each study the slopes will be intercorrelated.] An overview of the use of GLS for dependent standardized-mean-difference effect sizes was given by Raudenbush, Becker and Kalaian (1988), and we apply a similar approach here to sets of slopes.

To use GLS, we need estimates of the $P+1$ slopes from each of the $k$ samples (this includes the intercept $b_{i0}$) and their covariance matrices $\text{Cov}(\mathbf{b}_i)$. It is also possible to include studies that examine subsets of the $P$ predictors; we comment on how this would be done as we discuss details of the approach. Within sample $i$, the OLS estimate of $\boldsymbol{\beta}_i = (\beta_{i0}, \beta_{i1}, \ldots, \beta_{iP})$ is frequently reported. The estimator is

$$\mathbf{b}_i = (b_{i0}, b_{i1}, \ldots, b_{iP}) = (\mathbf{X}_i'\mathbf{X}_i)^{-1}\mathbf{X}_i'\mathbf{Y}_i,$$

with $\boldsymbol{\Sigma}_i = \text{Cov}(\mathbf{b}_i) = (\mathbf{X}_i'\mathbf{X}_i)^{-1}\sigma_i^2$, where $\mathbf{X}_i$ is the matrix of predictor values in the $i$th sample, plus a constant if the intercept is included. Typically $\sigma_i^2$ is not known, but rather is estimated with $S_i^2$, the mean squared error (MSE) of the regression in study $i$. In large samples, $\mathbf{b}_i$ is normally distributed with mean $\boldsymbol{\beta}_i$ and variance $\boldsymbol{\Sigma}_i$, which is the basis for the GLS approach. We will assume a common fixed-effects model (see Hedges and Vevea (1998)) which presumes that all samples incorporate the same $P$ predictors in the within-study regression model, and also assume that the vectors $\mathbf{b}_i$ estimate a common population slope vector $\boldsymbol{\beta}$.

We stack the $k$ sample slope vectors and make a blockwise diagonal matrix of the $\text{Cov}(\mathbf{b}_i)$ matrices, then apply GLS estimation. First define

$$\mathbf{b} = \begin{bmatrix} \mathbf{b}_1 \\ \mathbf{b}_2 \\ \vdots \\ \mathbf{b}_k \end{bmatrix}$$

and

$$\boldsymbol{\Sigma} = \begin{bmatrix} \text{Cov}(\mathbf{b}_1) & 0 & 0 & 0 \\ 0 & \text{Cov}(\mathbf{b}_2) & 0 & 0 \\ 0 & 0 & \cdots & 0 \\ 0 & 0 & 0 & \text{Cov}(\mathbf{b}_k) \end{bmatrix}.$$

Olkin (2003) pointed out that in some cases the covariance matrices $\text{Cov}(\mathbf{b}_i)$ could be pooled; below we discuss the case where a pooled MSE is available.

Then under the assumption that each slope vector $\mathbf{b}_i$ is estimating $\boldsymbol{\beta}$, we have the model

$$\mathbf{b} = \begin{bmatrix} b_{10} \\ b_{11} \\ \vdots \\ b_{1P} \\ \vdots \\ b_{k0} \\ \vdots \\ b_{kP} \end{bmatrix} = \mathbf{W}\boldsymbol{\beta} + \mathbf{e} = \begin{bmatrix} 1 & 0 & 0 & 0 \\ 0 & 1 & 0 & 0 \\ 0 & 0 & \cdots & 0 \\ 0 & 0 & 0 & 1 \\ \hline \vdots & \vdots & \vdots & \vdots \\ \hline 1 & 0 & 0 & 0 \\ 0 & 1 & \cdots & 0 \\ 0 & 0 & 0 & 1 \end{bmatrix} * \begin{bmatrix} \beta_0 \\ \beta_1 \\ \vdots \\ \beta_P \end{bmatrix} + \mathbf{e}.$$

The slopes are modeled as a function of $\boldsymbol{\beta}$ (the vector of $P+1$ population slopes) and a design matrix $\mathbf{W}$ composed of zeros and ones that identify which slopes are estimated in each sample. When all samples examine the same predictors, a stack of $(P+1) \times (P+1)$ identity matrices serves as $\mathbf{W}$ in the model $\mathbf{b} = \mathbf{W}\boldsymbol{\beta} + \mathbf{e}$, with $\text{Cov}(\mathbf{e}) = \text{Cov}(\mathbf{b})$ from above. If the samples do not all estimate the same model (i.e., some models use fewer than the full set of $P$ predictors), we can still use the GLS formulation to include those results in the synthesis. In such cases the component of $\mathbf{W}$ that represents a sample with fewer than $P$ predictors would not be a full identity matrix; row $p+1$ of the identity matrix for sample $i$ would be omitted if the $p$th predictor was not included in study $i$. However, as mentioned above, the interpretations (and distributions) of slopes from reduced models would not be exactly the same as for slopes from models with all $P$ predictors, and estimation of such quantities as $\sigma_i^2$ will be more complicated because $\sigma_i^2$ and $\sigma_{i'}^2$ (from samples $i$ and $i'$) may represent different population quantities if different sets of predictors were examined in samples $i$ and $i'$.

It is also possible to modify this approach somewhat to examine the influence of particular additional predictors on a focal predictor's slope. For instance, suppose the focus was on the role of teacher verbal ability as a predictor of student achievement. It could be of interest to see whether the slope for teacher verbal ability is different when a measure of students' prior achievement is included in the model. This could be accomplished by adding a column to the $\mathbf{W}$ matrix that would contain a 1 in each row representing a verbal-ability slope that came from a model that also included prior achievement. A small example illustrates this idea. Suppose that $X_1$ is teacher verbal ability, $X_2$ is prior achievement and $X_3$ is socioeconomic status. Two studies are available, only one of which (say study 1) includes prior achievement. The GLS model would be

$$\mathbf{b} = \begin{bmatrix} b_{10} \\ b_{11} \\ b_{12} \\ b_{13} \\ --- \\ b_{20} \\ b_{21} \\ b_{23} \end{bmatrix} = \mathbf{W}\boldsymbol{\beta} + \mathbf{e}$$

$$= \begin{bmatrix} 1 & 0 & 0 & 0 & 0 \\ 0 & 1 & 0 & 0 & 1 \\ 0 & 0 & 1 & 0 & 0 \\ 0 & 0 & 0 & 1 & 0 \\ - & - & - & - & - \\ 1 & 0 & 0 & 0 & 0 \\ 0 & 1 & 0 & 0 & 0 \\ 0 & 0 & 0 & 1 & 0 \end{bmatrix} * \begin{bmatrix} \beta_0 \\ \beta_1 \\ \beta_2 \\ \beta_3 \\ \gamma_1 \end{bmatrix} + \mathbf{e}.$$

The last column of $\mathbf{W}$ contains a 1 in row 2 (the row for the verbal-ability slope in study 1), showing that the first study included prior achievement ($X_2$) in the study-level model. Row 6—the row for the verbal-ability slope in study 2—does not have a 1 in the last column because study 2 did not include $X_2$. Also study 2 does not have a row of $\mathbf{W}$ with a 1 in column 2 because there is no estimate of $\beta_{22}$.

This model contains a fifth parameter, denoted $\gamma_1$ in the display, that represents the difference in the slope of teacher verbal ability when prior achievement is controlled. From this model we can determine that for study 1, the expected value of $b_{11}$ is $\beta_1 + \gamma_1$, while for study 2, $E[b_{21}] = \beta_1$. By including additional columns for key control variables or for other study features, the meta-analyst can examine hypotheses about whether the focal slope is affected by those elements of the original studies and their regression models. The details of such tests are described below. We caution, however, that including large numbers of added predictors may lead to multicollinearity, thus meta-analysts would be wise to carefully examine their models for the presence of this problem.

Regardless of the components of $\mathbf{W}$ and $\boldsymbol{\beta}$, we estimate $\boldsymbol{\beta}$ and its covariance as

$$\hat{\boldsymbol{\beta}}^* = (\mathbf{W}'\boldsymbol{\Sigma}^{-1}\mathbf{W})^{-1}\mathbf{W}'\boldsymbol{\Sigma}^{-1}\mathbf{b}$$

and

$$\text{Cov}(\hat{\boldsymbol{\beta}}^*) = (\mathbf{W}'\boldsymbol{\Sigma}^{-1}\mathbf{W})^{-1}.$$

Often, as noted above, we do not know $\text{Cov}(\mathbf{b}) = \boldsymbol{\Sigma}$; thus we typically substitute an estimate, which we shall denote $\mathbf{V}$, and compute instead

(4) $$\hat{\boldsymbol{\beta}} = (\mathbf{W}'\mathbf{V}^{-1}\mathbf{W})^{-1}\mathbf{W}'\mathbf{V}^{-1}\mathbf{b}$$

and

(5) $$\text{Cov}(\hat{\boldsymbol{\beta}}) = (\mathbf{W}'\mathbf{V}^{-1}\mathbf{W})^{-1}.$$

With large samples and under typical regularity conditions,

$$\hat{\boldsymbol{\beta}} \sim \text{N}(\boldsymbol{\beta}, \mathbf{Cov}(\hat{\boldsymbol{\beta}}));$$

thus confidence intervals for each element of $\boldsymbol{\beta}$ are available, using $\hat{\beta}_p \pm Z_{1-\alpha/2}\sqrt{C_{pp}}$, where $Z_{1-\alpha/2}$ is the upper tail $1-\alpha/2$ critical value of the standard normal distribution and $C_{pp}$ is the $p$th diagonal element of the $\text{Cov}(\hat{\boldsymbol{\beta}})$ matrix, the variance of $\hat{\boldsymbol{\beta}}$. Also a test of the hypothesis that the $p$th slope $\beta_p = 0$ can be obtained via

$$Z = \frac{\hat{\beta}_p}{\sqrt{C_{pp}}},$$

which is a standard normal deviate under the null hypothesis that $\beta_p = 0$. The value of $Z$ is compared to the cutpoints of the standard normal distribution.

Several other tests are available as well. A test of model fit, which is essentially a test of homogeneity of the regression intercepts and slopes across samples and across predictors, is given by

$$Q_{\text{E}} = (\mathbf{b} - \mathbf{W}\hat{\boldsymbol{\beta}})'\mathbf{V}^{-1}(\mathbf{b} - \mathbf{W}\hat{\boldsymbol{\beta}}),$$

which has a large-sample chi-squared distribution with $(k-1)(P+1)$ degrees of freedom if all slopes and intercepts are included. If $F$ additional columns are added to $\mathbf{W}$ to represent study features, then the degrees of freedom will be $(k-1)(P+F+1)$. If the magnitudes of the intercepts are not of interest, a modified $Q_{\text{E}}$ test can also be computed by including only the predictor slopes, thus reducing the dimension of $\mathbf{W}$ and including only those values of interest in $\mathbf{b}, \hat{\boldsymbol{\beta}}$ and $\mathbf{V}^{-1}$. In that case, $Q_{\text{E}}$ is chi-squared with $(k-1)P$ degrees of freedom. If $Q_{\text{E}}$ is large relative to cutpoints of the appropriate chi-squared distribution, the slopes vary beyond what one would expect to see given only sampling variability.

A test of the composite hypothesis that $\boldsymbol{\beta} = \mathbf{0}$ is given by

$$Q_B = \hat{\boldsymbol{\beta}}' \mathrm{Cov}(\hat{\boldsymbol{\beta}}) \hat{\boldsymbol{\beta}},$$

which is chi-squared with $P+1$ degrees of freedom under the null hypothesis that $\boldsymbol{\beta} = \mathbf{0}$ or with $P$ degrees of freedom if only predictor slopes are included (see, e.g., Hedges and Olkin (1985)).

### 4.1 Special Cases of the GLS Approach

The problem with the approach just described its that it is extremely rare to find the full covariance matrix of the slopes $\mathrm{Cov}(\mathbf{b})$ reported in a primary research study. Thus it is useful to note that the estimator shown in (4) simplifies to the weighted least squares (WLS) univariate estimator given in (2) if the off-diagonals in $\mathrm{Cov}(\mathbf{b})$ or $\mathbf{V}$ are set equal to zero.

Another special case is one in which it is possible to pool the estimates of $\sigma_i^2$ across studies. If all studies examine the same model and separate estimates of $\sigma_i^2$ are available, then it is possible to remove the MSE values from the $\mathrm{Cov}(\mathbf{b})$ matrices and use a blockwise diagonal matrix $\mathbf{X}^*$ containing the $(\mathbf{X}_i' \mathbf{X}_i)^{-1}$ matrices in place of $\mathbf{V}$ in formulas (4) and (5). It is shown in the Appendix that

(6) $\quad \hat{\boldsymbol{\beta}} = (\mathbf{W}' (\mathbf{X}^*)^{-1} \mathbf{W})^{-1} \mathbf{W}' (\mathbf{X}^*)^{-1} \mathbf{b}$

produces an estimate of $\boldsymbol{\beta}$ equivalent to the value that would be obtained from a pooled sample. This is because $(\mathbf{X}^*)^{-1}$ is a blockwise diagonal matrix containing the values of $\mathbf{X}_i' \mathbf{X}_i$, and the product $\mathbf{W}' (\mathbf{X}^*)^{-1} \mathbf{W}$ sums the $\mathbf{X}_i' \mathbf{X}_i$ values across the $k$ studies. Similarly $(\mathbf{X}^*)^{-1} \mathbf{b}$ equals the sum across studies of the values of the products $\mathbf{X}_i' \mathbf{Y}_i$, leading to equivalence with the estimator based on the pooled sample.

Values of $(\mathbf{X}_i' \mathbf{X}_i)^{-1}$ can be estimated if each study reports the covariance matrix for the slopes and $S_i^2$, the estimate of $\sigma_i^2$ (or other quantities that allow computation of $S_i^2$, e.g., the variance of the outcome and the $R^2$ for the regression). Each element of $\mathrm{Cov}(\mathbf{b})$ is divided by the estimated MSE: $(\mathbf{X}_i' \mathbf{X}_i)^{-1} = \mathrm{Cov}(\mathbf{b}_i) S_i^{-2}$. This method requires that a pooled value of the MSE (say $S_*^2$) be obtained and multiplied by the covariance of the synthesized slope estimator computed using $\mathbf{X}^*$, to compensate for $S_i^2$ being removed when $\mathbf{X}^*$ is substituted for $\mathbf{V}$. Thus the matrix of covariances among the synthesized slopes is

(7) $\quad \mathrm{Cov}(\hat{\boldsymbol{\beta}}) = (\mathbf{W}' (\mathbf{X}^*)^{-1} \mathbf{W})^{-1} S_*^2.$

One possible estimator $S_*^2$ could be

$$S_*^2 = \sum_i dfe_i S_i^2 \Big/ \sum_i dfe_i,$$

where $dfe_i$ is the degrees of freedom for error in study $i$. Unfortunately primary researchers do not always report the value of $S_i^2$, the mean squared error of the regression model in the primary study. Given this and the rarity of finding full $\mathrm{Cov}(\mathbf{b}_i)$ matrices, it is expected that this special case will be relatively uncommon.

### 4.2 Limitations

The discussion of special cases focuses our attention to the fact that one weakness of the proposed GLS approach is that it uses the $\mathrm{Cov}(\mathbf{b}_i)$ matrices that are rarely reported. It is unlikely, even with more stringent reporting requirements, that authors will routinely begin to report these matrices, particularly in primary research studies where many models with large numbers of predictors are estimated and compared.

There are two possible approaches to this problem. One is to simply assume the slopes are independent, use the squared standard errors of the slopes as the diagonal elements of $\mathrm{Cov}(\mathbf{b}_i)$ and set the off-diagonal elements to zero. This produces weighted least squares estimates. A slightly more conservative approach would be to assume a common correlation value among all slopes [e.g., $\mathrm{Corr}(b_{ip}, b_{ip'}) = 0.2$] and then compute the off-diagonal elements of each $\mathrm{Cov}(\mathbf{b}_i)$ matrix as the product of the slope standard errors (SEs) and that common correlation, specifically, $\mathrm{Cov}(b_{ip}, b_{ip'}) = \mathrm{Corr}(b_{ip}, b_{ip'})^* \mathrm{SE}(b_{ip})^* \mathrm{SE}(b_{ip'})$.

One final point regarding this issue relates to model specification in the primary research studies in the meta-analysis. That is, if a model is well specified in study $i$, there should be no serious multicollinearity and the degree of covariation among the slopes in $\mathrm{Cov}(\mathbf{b}_i)$ should not be great. In such cases, setting all off-diagonal elements of $\mathrm{Cov}(\mathbf{b}_i)$ to zero would not have serious consequences. However, reporting conventions in many fields do not require authors to mention whether multicollinearity was assessed or to report on multicollinearity diagnostics. So the meta-analyst must trust that the primary study authors actually checked for multicollinearity and that any models reported upon are relatively free from this problem.

## 5. EXAMPLE

In this example we use data from the base year of the National Education Longitudinal Study of 1988 (NELS:88). NELS:88 is a survey of a national sample of high-school students from over 1000 schools. The same measures are used across schools and when analyzed with proper weights, the full sample represents the U.S. grade 10 high-school population from 1988. In our example we use as "studies" 13 schools with samples of more than 45 students; we do not use the NELS:88 sampling weights that would produce results that reflect the national population. Both the school-level sampling weights and the within-school weights that could be used to make each school's estimates reflective of the population of that school were ignored.

### 5.1 Model

Our regression model uses three of the standardized cognitive tests administered as part of the NELS:88 survey—the science, mathematics and reading scales. This model views science achievement as a function of math and reading test performance. Specifically, $Y$ represents the NELS:88 science achievement test, $X_1$ is the mathematics test and $X_2$ is the reading test. The model estimated in study $i$ is

$$Y_{ij} = \beta_0 + \beta_1 X_{1j} + \beta_2 X_{2j} + e_{ij}$$

for student $j$, with error $e_{ij}$. We use ordinary least squares to obtain school level estimates of this model. Computations were done using PROC REG and PROC IML in SAS.

Our analyses are based on the item-response-theory estimated number-right scores for these test batteries; therefore the raw slopes can be interpreted as the predicted change in the science test score for a one item increase in the math or reading test score. The science test had 25 items, the math test had 40 and the reading test had 21 items. Means across the 13 schools were 13.5 or 54% correct on science ($SD = 5.7$), 24.4 or 61% correct on math ($SD = 10.4$) and 13.9 or 66% correct for reading ($SD = 5.7$). The correlation between math and reading scores was $r_{MR} = 0.70$, and each predictor was also correlated with the outcome at about that same level ($r_{MS} = 0.70$, $r_{RS} = 0.67$) in the full sample.

### 5.2 Results

The regression model with $X_1$ and $X_2$ as predictors of $Y$ was estimated within each of the schools, and the slope estimates and fitted models are shown in Table 1. The data from the 13 schools were also pooled (used as a single sample) and the full model including intercepts was estimated across all schools (for all 664 cases); this result is labeled "Full sample." The estimated model from this analysis of the 13 schools together was $\hat{Y}_j = 2.552 + 0.245X_{1j} + 0.358X_{2j}$ and it is shown in the first row of Table 1. (The subscript $j$ has been omitted from the table entries for simplicity.) Inspection of the models for the 13 schools shows some variation in the slopes and intercepts; the most unusual looking model is for school 3. Also casual inspection of the mean squared errors shows some variation in the $S_i^2$ values, with school 10 showing the smallest value. However, Levene's test suggests the error variances are not different $[F(12, 651) = 1.25, p = 0.25]$, indicating that it is reasonable to proceed with the analysis based on the pooled MSE. (Although here we have the raw data and can compute Levene's test, in practice other tests that do not require raw data such as $F_{\max}$ or Cochran's $C$ could be used to test residual variance equality.)

TABLE 1
*Fitted regressions and MSE values for full sample and 13 schools*

| Sample | $n_i$ | Fitted regression | MSE ($S_i^2$ for school$_i$) |
|---|---|---|---|
| Full | 664 | $2.552 + 0.245X_1 + 0.358X_2$ | 14.44 |
| 1 | 64 | $5.470 + 0.219X_1 + 0.260X_2$ | 17.46 |
| 2 | 59 | $3.591 + 0.246X_1 + 0.270X_2$ | 14.24 |
| 3 | 67 | $5.619 + 0.040X_1 + 0.638X_2$ | 14.05 |
| 4 | 45 | $4.381 + 0.181X_1 + 0.392X_2$ | 10.75 |
| 5 | 47 | $4.305 + 0.260X_1 + 0.282X_2$ | 9.32 |
| 6 | 45 | $2.346 + 0.185X_1 + 0.195X_2$ | 14.60 |
| 7 | 45 | $0.228 + 0.283X_1 + 0.339X_2$ | 9.80 |
| 8 | 56 | $2.289 + 0.289X_1 + 0.312X_2$ | 13.32 |
| 9 | 45 | $3.600 + 0.248X_1 + 0.263X_2$ | 12.65 |
| 10 | 51 | $2.156 + 0.192X_1 + 0.498X_2$ | 6.50 |
| 11 | 48 | $3.621 + 0.133X_1 + 0.413X_2$ | 11.02 |
| 12 | 45 | $3.144 + 0.250X_1 + 0.382X_2$ | 17.65 |
| 13 | 47 | $3.781 + 0.251X_1 + 0.151X_2$ | 13.20 |

The upper triangles of the covariance, correlation and $\mathbf{X}_i'\mathbf{X}_i$ matrices among the slopes for three of the schools and the full sample are shown in Table 2. The $\mathbf{X}_i'\mathbf{X}_i$ matrices are used in the third method of estimation using the pooled MSE. The elements of the $\text{Cov}(\mathbf{b}_i)$ matrices are obtained as the products of the entries in $(\mathbf{X}_i'\mathbf{X}_i)^{-1}$ times the MSE [e.g., for school 1, the first entry in $\text{Cov}(\mathbf{b}_1)$ is 1.934, which is within rounding error of $17.463 \times 0.1107 = 1.933$]. Also the MSE pooled across the 13 schools is $S_*^2 = 12.83$.

Table 3 repeats the OLS results for the pooled sample (to facilitate comparisons) and also presents the slopes estimated using the three synthesis methods described above. The first set of results is based on the GLS estimation method with mean and variance given in (4) and (5). While the intercept differs somewhat from the pooled-sample intercept, the slope coefficients are both within 0.015 of the values estimated in the full sample. Considering that the slopes represent predicted change on a 25-point science test (given a one-point change on $X$), these are very small differences. The test of homogeneity of the models using $Q_E$ defined above for all slopes and intercepts shows that indeed the slopes and intercepts are not homogeneous ($Q_E = 114.16$, df $= 36$, $p < 0.001$), and may not have come from a single population. However, this test asks whether all parameters are equal across schools; thus the test can also be large if the intercepts differ. The test can be computed for the predictor slopes only (omitting $b_0$ values): when this is done, the $Q_E$ value is smaller ($Q_E = 21.74$, df $= 24$, $p = 0.59$), and indicates the math and reading slopes are homogeneous across schools. Also at least one of the slopes differs from zero, according to the $Q_B$ test ($Q_B = 518.16$, df $= 2$, $p < 0.001$).

TABLE 2
*Covariance and $\mathbf{X'X}$ matrices for three studies and full sample*

| Sample | | | $\mathbf{X'X}$ | | | Cov (b) | | | Corr (b) | |
|---|---|---|---|---|---|---|---|---|---|---|
| | | I | M | R | I | M | R | | M | R |
| Full | I | 0.01175 | −0.00018 | −0.00042 | 0.1697 | −0.0026 | −0.0060 | | −0.32 | −0.40 |
| ($n = 664$) | M | | 0.00003 | −0.00003 | | 0.0014 | −0.0005 | | | −0.70 |
| | R | | | 0.00009 | | | 0.0013 | | | |
| School 1 | I | 0.1107 | −0.0037 | −0.0017 | 1.9340 | −0.0648 | −0.0302 | | −0.61 | −0.22 |
| ($n_1 = 64$) | M | | 0.0003 | −0.0002 | | 0.0058 | −0.0043 | | | −0.57 |
| | R | | | 0.0006 | | | 0.0098 | | | |
| School 2 | I | 0.0914 | −0.0015 | −0.0034 | 1.3018 | −0.0218 | −0.0482 | | −0.36 | −0.44 |
| ($n_2 = 59$) | M | | 0.0002 | −0.0002 | | 0.0028 | −0.0030 | | | −0.60 |
| | R | | | 0.0006 | | | 0.0092 | | | |
| School 3 | I | 0.4267 | −0.0103 | −0.0058 | 5.9953 | −0.1449 | −0.0817 | | −0.65 | −0.26 |
| ($n_3 = 67$) | M | | 0.0006 | −0.0004 | | 0.0082 | −0.0063 | | | −0.54 |
| | R | | | 0.0012 | | | 0.0164 | | | |

TABLE 3
*Results of synthesis*

| Method of estimation | Slope estimates | | | | Cov (b) | | | | Corr (b) | |
|---|---|---|---|---|---|---|---|---|---|---|
| | Intercept | Math | Reading | | I | M | R | | | |
| Full sample | 2.552 | 0.245 | 0.358 | I | 0.1697 | −0.0026 | −0.0060 | | −0.32 | −0.40 |
| ($n = 664$) | | | | M | | 0.0004 | −0.0005 | | | −0.70 |
| | | | | R | | | 0.0013 | | | |
| GLS | 2.268 | 0.247 | 0.373 | I | 0.1463 | −0.0021 | −0.0054 | | −0.30 | −0.41 |
| | | | | M | | 0.0003 | −0.0004 | | | −0.71 |
| | | | | R | | | 0.0012 | | | |
| WLS | 2.936 | 0.221 | 0.343 | I | 0.1747 | 0 | 0 | | 0 | 0 |
| | | | | M | | 0.0004 | 0 | | | 0 |
| | | | | R | | | 0.0012 | | | |
| GLS using $(\mathbf{X'X})^{-1}$ | 2.552 | 0.245 | 0.358 | I | 0.1507 | −0.0023 | −0.0053 | | −0.32 | −0.40 |
| | | | | M | | 0.0004 | −0.0004 | | | −0.70 |
| | | | | R | | | 0.0012 | | | |

At this point more detailed analyses of slopes for each predictor might be of use, and standard univariate meta-analysis procedures (e.g., Hedges and Olkin (1985)) could be applied to each set of slope values, or other GLS based analyses can be used if it is desired to model the vectors of slopes (Raudenbush, Becker and Kalaian, 1988). Also to explore between-studies differences in models one could then examine moderating variables as described above. If the slopes or parameters for additional study features still did not appear homogeneous, one could estimate between-studies variance components for each of the slopes. A variety of estimators for the between-studies variance exist (e.g., Hedges and Olkin (1985); Sidik and Jonkman, 2005) and an estimated between-studies variance could then be added to each study's sampling variance to augment its uncertainty.

The next set of results was obtained by eliminating the off-diagonal elements from the Cov(**b**) matrices. This is equivalent to estimating the slopes using the univariate methods shown in displays (2) and (3). These values also do not deviate far from the full-sample values; both slopes are within 0.025 points of the slopes from the full sample—deviating only slightly more than the GLS values. This is in spite of the fact that the predictors and outcome show moderate intercorrelations as can be seen by inspection of the Corr(**b**) matrices shown in Table 2. Finally, the third set of results is computed using the $\mathbf{X}^*$ matrix in place of the Cov(**b**) matrix, and the pooled MSE in place of each $S_i^2$ value. As noted above the slope computed in this way is identical to the slope for the full sample, and the covariance matrix differs from the full sample matrix by a constant factor equal to the ratio of the estimated pooled MSE to the full sample MSE (here that ratio is $12.83/14.44 = 0.89$). It is somewhat problematic that the variances of slopes from the meta-analysis are less than or equal to the values from the full sample (thus suggesting more precision). From one application it is not possible to determine whether this is a result of the particular nature of the example data (11 of 13 schools show MSEs smaller than the MSE of 14.44 for the full data set) or something more pervasive. Further examination of the performance of these estimation methods via Monte Carlo methods will indicate whether a consistent pattern of underestimation is found.

Our new method takes into account the interrelationships among predictors from the primary studies, as well as heteroscedasticity of the slopes, via the variance–covariance matrix of the slopes. Both features should represent improvements on ordinary least squares methods. Such OLS approaches typically include dummy variables to show the presence of specific predictors or study features, but do not deal with the possible dependence of the predictors in the model(s), nor do they account for the heteroscedasticity inherent in the slope estimates. Even when off-diagonal elements of Cov(**b**) were set to zero in our analysis, the weighted least squares slopes were very close to the full sample slopes.

## 6. CONCLUSION

This paper presents a review of existing methods for the synthesis of regression slopes and a new multivariate approach based on generalized least squares estimation that is applicable to the meta-analytic context. Table 4 summarizes the main strengths and weaknesses of all of the methods. Two methods require raw data and thus are not appropriate for the meta-analysis context. Five others focus only on a single focal slope (or some related index such as a $t$ test of that slope) and thus cannot provide an overall model based on the synthesis. Also these methods ignore dependence among slopes by omitting all but the focal slope. Some additionally ignore the inherent differential precision of slopes across studies by applying ordinary least squares estimation methods. The new multivariate GLS method addresses these problems, but is itself limited because information about covariation among slopes is typically not given in primary research reports.

A comparison of the results of three variations of the GLS approach applied to an educational data set is made to the analysis of all data in a single pooled analysis. The analyses produce very similar results and in some cases have identical results (given the availability of specific summary statistics such as mean squared errors for the individual regression models). Our results emphasize the importance of full reporting of sufficient statistics in primary research studies; with less complete information, the full GLS analysis is not possible. However, even the less complex weighted least squares approach appeared to provide reasonable values in one example analysis.

## APPENDIX: EQUIVALENCE OF FULL SAMPLE AND SYNTHESIZED RESULTS WHEN $\sigma_i^2 = \sigma^2$ FOR $i = 1$ TO $k$

Consider $k$ independent samples or studies each examining a model relating predictors $X_1$ through

$X_P$ to an outcome $Y$ for case $j$. Specifically, in study $i$,

$$Y_{ij} = \beta_{i0} + \beta_{i1} X_{ij1} + \cdots + \beta_{iP} X_{ijP} + e_{ij}$$

for $j = 1$ to $n_i$.

For later use we also define $\mathbf{X}$ and $\mathbf{Y}$ by stacking the individual $\mathbf{X}_i$ and $\mathbf{Y}_i$ matrices:

$$\mathbf{X} = \begin{bmatrix} \mathbf{X}_1 \\ \mathbf{X}_2 \\ \vdots \\ \mathbf{X}_k \end{bmatrix} \quad \text{and} \quad \mathbf{Y} = \begin{bmatrix} \mathbf{Y}_1 \\ \mathbf{Y}_2 \\ \vdots \\ \mathbf{Y}_k \end{bmatrix}.$$

The OLS regression slope for the full combined sample is

(A1) $$\mathbf{b}^* = (\mathbf{X}'\mathbf{X})^{-1}\mathbf{X}'\mathbf{Y}.$$

Within study $i$, the OLS estimate of $\boldsymbol{\beta}_i = (\beta_{i0}, \beta_{i1}, \ldots, \beta_{iP})$ is

$$\mathbf{b}_i = (b_{i0}, b_{i1}, \ldots, b_{iP}) = (\mathbf{X}_i'\mathbf{X}_i)^{-1}\mathbf{X}_i'\mathbf{Y}_i$$

with

$$\boldsymbol{\Sigma}_i = \text{Cov}(\mathbf{b}_i) = (\mathbf{X}_i'\mathbf{X}_i)^{-1}\sigma_i^2.$$

If it is reasonable to assume that the error variances $\sigma_i^2$, for $i = 1$ to $k$ are equal (e.g., if the $k$ samples are drawn from one population), then we have

$$\boldsymbol{\Sigma}_i = \text{Cov}(\mathbf{b}_i) = (\mathbf{X}_i'\mathbf{X}_i)^{-1}\sigma^2.$$

Next we define

$$\mathbf{b} = \begin{bmatrix} \mathbf{b}_1 \\ \mathbf{b}_2 \\ \vdots \\ \mathbf{b}_k \end{bmatrix}$$

and

$$\boldsymbol{\Sigma} = \begin{bmatrix} \text{Cov}(\mathbf{b}_1) & 0 & 0 & 0 \\ 0 & \text{Cov}(\mathbf{b}_2) & 0 & 0 \\ 0 & 0 & \cdots & 0 \\ 0 & 0 & 0 & \text{Cov}(\mathbf{b}_k) \end{bmatrix}$$

$$= \begin{bmatrix} (\mathbf{X}_1'\mathbf{X}_1)^{-1} & 0 & 0 & 0 \\ 0 & (\mathbf{X}_2'\mathbf{X}_2)^{-1} & 0 & 0 \\ 0 & 0 & \cdots & 0 \\ 0 & 0 & 0 & (\mathbf{X}_k'\mathbf{X}_k)^{-1} \end{bmatrix} \sigma^2,$$

TABLE 4
*Methods of summarizing slopes*

| Method | Data needed | Strength | Weakness |
| --- | --- | --- | --- |
| Simple slope summaries | Slopes | Simple, little data needed | Focuses on a single focal slope; ignores dependence and precision of slopes |
| Summaries of $t$ statistics | $t$ values for slope tests | Simple; little data needed; $X$s and $Y$s can be on any scales | Focuses on only a single focal slope; $t$ values contain irrelevant information about sample size; unclear how an index of effect is obtained |
| Iterative least squares approach | Raw data | Accounts for covariation among predictors | Iteration needed to get covariance matrix |
| Dose–response models (WLS approach for dichotomous outcomes) | Slopes and standard errors for models with dichotomous outcomes | Weights by precision | Focuses on only a single focal slope; ignores dependence of slopes |
| Validity generalization approach | Slopes, reliabilities of $X$ and sample sizes | Simple; little data needed | Reliabilities often not reported; ignores dependence of slopes |
| Univariate WLS approach | Slopes and standard errors | Relatively simple; weights by precision | Focuses on only a single focal slope; ignores dependence of slopes |
| Multivariate Bayesian approach | Raw data | Collateral information can be shared across studies | Multistage formulation; requires priors and hyperparameters; $X$s and $Y$s must be on same scales |
| Multivariate GLS approach | Slopes and Cov($\mathbf{b}$) matrices | Weights by precision; accounts for covariation; provides entire pooled model | Requires covariances among slopes, which are often not reported |

THE SYNTHESIS OF REGRESSION SLOPES IN META-ANALYSIS 15which is labeled $\mathbf{X}^*\sigma^2$ in the text. When inverted, this matrix is

$$\boldsymbol{\Sigma}^{-1} = \begin{bmatrix} (\mathbf{X}_1'\mathbf{X}_1) & \mathbf{0} & \mathbf{0} & \mathbf{0} \\ \mathbf{0} & (\mathbf{X}_2'\mathbf{X}_2) & \mathbf{0} & \mathbf{0} \\ \mathbf{0} & \mathbf{0} & \cdots & \mathbf{0} \\ \mathbf{0} & \mathbf{0} & \mathbf{0} & (\mathbf{X}_k'\mathbf{X}_k) \end{bmatrix} \sigma^{-2}.$$

Also the $i$th cross-product matrix is

$$\mathbf{X}_i^*\mathbf{X}_i = \begin{bmatrix} n_i & \sum_j X_{ij1} & \sum_j X_{ij2} & \cdots & \sum_j X_{ijP} \\ \sum_j X_{ij1} & \sum_j X_{ij1}^2 & \sum_j X_{ij2}X_{ij1} & \cdots & \sum_j X_{ij1}X_{ijP} \\ \vdots & \vdots & \vdots & & \vdots \\ \sum_j X_{ijP} & \sum_j X_{ij1}X_{ijP} & \sum_j X_{ij2}X_{ijP} & \cdots & \sum_j X_{ijP}^2 \end{bmatrix}.$$

The synthesized GLS slope estimator is

(A2) $\qquad \hat{\boldsymbol{\beta}}^* = (\mathbf{W}'\boldsymbol{\Sigma}^{-1}\mathbf{W})^{-1}\mathbf{W}'\boldsymbol{\Sigma}^{-1}\mathbf{b},$

where $\mathbf{W}$ is a stack of identity matrices of dimension $P+1$. The first component of the estimator is $(\mathbf{W}'\boldsymbol{\Sigma}^{-1}\mathbf{W})$, which is a sum of matrices:

$$(\mathbf{W}'\boldsymbol{\Sigma}^{-1}\mathbf{W}) = (\mathbf{X}_1'\mathbf{X}_1)\sigma^2 + (\mathbf{X}_2'\mathbf{X}_2)\sigma^2 + \cdots + (\mathbf{X}_k'\mathbf{X}_k)\sigma^2.$$

Equivalently,

$$\mathbf{W}'\boldsymbol{\Sigma}^{-1}\mathbf{W} = \begin{bmatrix} \sum_i n_i & \sum_i \sum_j X_{ij1} & \sum_i \sum_j X_{ij2} & \cdots & \sum_i \sum_j X_{ijP} \\ \sum_i \sum_j X_{ij1} & \sum_i \sum_j X_{ij1}^2 & \sum_i \sum_j X_{ij2}X_{ij1} & \cdots & \sum_i \sum_j X_{ij1}X_{ijP} \\ \vdots & \vdots & \vdots & & \vdots \\ \sum_i \sum_j X_{ijP} & \sum_i \sum_j X_{ij1}X_{ijP} & \sum_i \sum_j X_{ij2}X_{ijP} & \cdots & \sum_i \sum_j X_{ijP}^2 \end{bmatrix} \sigma^2,$$

which is simply $\mathbf{X}'\mathbf{X}\sigma^2$ for the full sample (i.e., the sample pooled across studies), so $\mathbf{W}'\boldsymbol{\Sigma}^{-1}\mathbf{W} = (\mathbf{X}'\mathbf{X})^{-1}\sigma^{-2}$. Thus we can write

(A3) $\qquad \hat{\boldsymbol{\beta}}^* = [(\mathbf{X}'\mathbf{X})^{-1}\sigma^{-2}]\mathbf{W}'\boldsymbol{\Sigma}^{-1}\mathbf{b}.$

Next we consider the term $\mathbf{W}'\boldsymbol{\Sigma}^{-1}\mathbf{b}$. The product $\mathbf{W}'\boldsymbol{\Sigma}^{-1}$ is a matrix that is $\sigma^2$ times a concatenation of $(\mathbf{X}_i'\mathbf{X}_i)$ matrices, specifically

$$\mathbf{W}'\boldsymbol{\Sigma}^{-1} = [\mathbf{X}_1'\mathbf{X}_1|\mathbf{X}_2'\mathbf{X}_2|\cdots \mathbf{X}_i'\mathbf{X}_i \cdots |\mathbf{X}_k'\mathbf{X}_k]\sigma^2.$$

Also, $\mathbf{b}$ is the stacked vector of the $k$ individual sample slope vectors. Thus

$$\mathbf{W}'\boldsymbol{\Sigma}^{-1}\mathbf{b} = \mathbf{X}_1'\mathbf{X}_1\mathbf{b}_1\sigma^2 + \mathbf{X}_2'\mathbf{X}_2\mathbf{b}_2\sigma^2 + \cdots + \mathbf{X}_k'\mathbf{X}_k\mathbf{b}_k\sigma^2.$$

Each component of this sum is a $(P+1) \times (P+1)$ matrix. Then substituting $\mathbf{b}_i = (\mathbf{X}_i'\mathbf{X}_i)^{-1}\mathbf{X}_i'\mathbf{Y}_i$ into this equation, we obtain

$$\mathbf{W}'\boldsymbol{\Sigma}^{-1}\mathbf{b} = \mathbf{X}_1'\mathbf{X}_1(\mathbf{X}_1'\mathbf{X}_1)^{-1}\mathbf{X}_1'\mathbf{Y}_1\sigma^2 + \cdots + \mathbf{X}_k'\mathbf{X}_k(\mathbf{X}_k'\mathbf{X}_k)^{-1}\mathbf{X}_k'\mathbf{Y}_k\sigma^2$$
$$= \sigma^2\left[\sum \mathbf{X}_i'\mathbf{Y}_i\right] = \sigma^2 \mathbf{X}'\mathbf{Y}.$$

Substituting this result into (A3), we see that

$$\hat{\boldsymbol{\beta}}^* = (\mathbf{X}'\mathbf{X})^{-1}\sigma^{-2}\mathbf{W}'\boldsymbol{\Sigma}^{-1}\mathbf{b} = (\mathbf{X}'\mathbf{X})^{-1}\sigma^{-2}[\sigma^2 \mathbf{X}'\mathbf{Y}]$$
$$= (\mathbf{X}'\mathbf{X})^{-1}\mathbf{X}'\mathbf{Y},$$

which equals $\mathbf{b}^*$ given in (A1).

## ACKNOWLEDGMENTS

This work was supported by NSF Grants REC-0335656 and REC-0634013. Parts of this work were presented in the symposium "Multivariate Analysis: In Celebration of Ingram Olkin's 80th Birthday" at the Joint Statistical Meetings, Toronto, Canada, August 2004.## REFERENCES

AMEMIYA, Y. and FULLER, W. A. (1984). Estimation for the multivariate errors-in-variables model with estimated error covariance matrix. *Ann. Statist.* **12** 497–509. MR0740908

ASHENFELTER, O., HARMON, C. and OOSTERBEEK, H. (1999). A review of estimates of the schooling/earnings relationship, with tests for publication bias. *Labour Economics* **6** 453–470.

BAKER, C. B., TWEEDIE, R., DUVAL, S. and WOODS, S. W. (2003). Evidence that the SSRI dose response in treating major depression should be reassessed: A meta-analysis. *Depression and Anxiety* **17** 1–9.